\documentstyle[12pt]{article}
\begin{document}
\begin{titlepage}
\title{The influence of structure disorder on mean atomic momentum
fluctuations and a spin-wave spectrum
}
\author{I. Vakarchuk, V. Tkachuk, T. Kuliy
\\[1.5ex]
{\small Theoretical Physics Chair, Ivan Franko Lviv State University}
\\ {\small 12 Drahomanov St., L'viv, 290005, Ukraine}
\\ {\small phone: +7-0322-728080, fax: +7-0322-727981,
e-mail: taras@Big.LITech.Lviv.UA}
}
\maketitle\thispagestyle{empty}
\begin{abstract}
The relation between atomic momenta fluctuations and density
fluctuations is obtained in frames of mean-field approximation.
Using two-time temperature Green functions within Tyablikov approximation
the equations for spin excitation energy and damping are obtained.
The asymptotics of energy and damping in the long-wave limit are investigated
and the anomalous behaviour of spin-wave stiffness constant is discussed.
\end{abstract}
\end{titlepage}

\renewcommand{\theequation}{\arabic{equation}}
\bigskip \centerline{ \Large \bf Introduction} \bigskip

In amorphous ferromagnets magnetic momenta of atoms do not have fixed values.
Structure disorder leads to the differences in the mean momentum $<S^z_i>$
of atoms localized at different sites $i$. The first attempt
to take into account the structural fluctuations of atomic momenta in order to
explain the anomalous behaviour of spin-wave stiffness constant in amorphous
ferromagnets, increasing with increasing external field, was done in ref.
\cite{Kan}.
Kaneyoshi in \cite{Kan} assumes that the fluctuations of each momenta are
statistically independent of each other and their values are given by the
Gauss distribution function. Note, that the atomic momenta fluctuations are
entirely connected to the atomic density fluctuations.

The aim of this paper is to determine the relations between the atomic density
fluctuations and the atomic momenta fluctuations and to investigate the
influence of atomic momenta structural fluctuation on the spectrum and damping
of spin waves in amorphous ferromagnets.

\bigskip\bigskip\bigskip
\renewcommand{\theequation}{1.\arabic{equation}}
\centerline{ \Large \bf 1. Green Functions of the structurally Disordered}
\bigskip
\centerline{ \Large \bf Heisenberg Model.} \bigskip

Let us consider a structurally disordered system of $N$ spins in the volume
$V$ which is described by the isotropic Heisenberg Hamiltonian

\begin{equation}  \label{1}
    H = -{1 \over 2} \sum_{i,j} J_{ij} {\bf S}_i {\bf S}_j -h \sum_{j}  S^z_j,
\end{equation}
where $J_{ij}=J(|{\bf R}_i-{\bf R}_j|)$ is the exchange integral describing
interaction
between the $i-$th and $j-$th atoms, ${\bf S}_i$ is the spin operator
of the $i-$th atom, $h$ is external magnetic field, ${\bf R}_i$ are the
coordinates of atoms, that are randomly distributed.

We use the two-time temperature Green function method for the investigation of
the spin excitations. The Green function within Tyablikov approximation
satisfies the equation of motion

\begin{equation}  \label{2} 
(E-h)\ll S^+_l|S^-_{l'}\gg=2\delta_{l,l'}\sigma_l +\sum_{j(\not=l)} J_{ij}
\Theta(R_{ij}-a)\ll \sigma_j S^+_l - \sigma_l S^+_j|S^-_{l'}\gg ,
\end{equation}
where $S^\pm _l=S_x\pm i S_y$, $\sigma_l=<S^z_l>$ We introduce
the $\Theta-$ function in
the equation (\ref{2}) to take into account during the operation of
configurational
averaging the fact that the minimum distance between atoms is $a$ \cite{Vak}.
Traditionally, the second approximation is specific for disordered systems and
means the neglecting of structural fluctuations of spin momentum
$\sigma_l \rightarrow \overline\sigma_l=\sigma$
where symbol $\overline{(\dots)}$ denotes the random configurational
ensemble average. In our paper we do not perform this approximation and
therefore we have possibility to take into account the fluctuations of spin.

The spin excitation spectrum can be obtained from the configurationally
averaged Green function $\overline{\ll S^+_q|S^-_{q'}\gg }$,
where
$S_{\bf q}^\pm=
{1\over\sqrt{N}} \sum\limits_{j=1}^N S^\pm_j e^{-i {\bf q}{\bf R}_j}$.

For the Green function in the momentum space on the basis of (\ref{2}) we
obtain the equation

\begin{eqnarray}  \label{3} 
(E-E_0(q))\ll S^+_{\bf q}|S^-_{\bf q'}\gg =2\sigma\delta({\bf q}+{\bf q'}) +
2{1\over\sqrt{N}} \Delta\sigma_{{\bf q}+{\bf q'}} + \\
 + {N\over V}\sum_{\bf k}(J(|{\bf q}-{\bf k}|)-J(k)){1\over\sqrt{N}}
\Delta\sigma_{{\bf q}-{\bf k}}\ll S^+_{\bf k}|S^-_{\bf q'}\gg , \nonumber
\end{eqnarray}
where	$\Delta\sigma_{\bf q}=\sigma_{\bf q}-\sigma\sqrt{N}\delta({\bf q})$,
\ \
$\sigma_{\bf q}= {1\over\sqrt{N}}
\sum\limits_{j=1}^N \sigma_j e^{-i {\bf q}{\bf R}_j}$, \ \
$E_0(q)=\sigma {N\over V}(J(0)-J(q))$, \\
$J(q)=\int {d{\bf R} \Theta (R-a) J(R) e^{-i {\bf q}{\bf R}}}$,\ \
$\delta ({\bf q})$ \
is the Kronecker symbol.

Averaging (\ref{3}) over possible realization of atomic configurations,
the equation for averaged Green function can be written in the following form:

\begin{eqnarray}  \label{4} 
(E-E_0(q))\overline{\ll S^+_{\bf q}|S^-_{\bf q'}\gg }=2\sigma\delta({\bf q}+
{\bf q'}) + \\
 + {N\over V}\sum_{\bf k}(J(|{\bf q}-{\bf k}|)-J(k)){1\over\sqrt{N}}
\overline{\Delta\sigma_{{\bf q}-{\bf k}}\ll S^+_{\bf k}|S^-_{\bf q'}\gg }.
\nonumber
\end{eqnarray}

Equation (\ref{4}) contains a higher-order averaged Green function
$\overline{\Delta\sigma G}$. One can write the equation for this function,
multiplying $\Delta\sigma_{\bf k'}$ by (\ref{3}) \medskip
and performing configurational
averaging. These equations include $\overline{\Delta\sigma\Delta\sigma G}$.
To solve these equations the decoupling of configurational
averages is used
\[
\overline{\Delta\sigma_{{\bf q}-{\bf k}}\Delta\sigma_{{\bf k}-{\bf k'}}\ll
S^+_{\bf k'}|S^-_{\bf q'}\gg }
\approx\overline{\Delta\sigma_{{\bf q}-{\bf k}}
\Delta\sigma_{{\bf k}-{\bf k'}}}
\times\overline{\ll S^+_{\bf k'}|S^-_{\bf q'}\gg },
\]
where~~ $\overline{\Delta\sigma_{{\bf q}-{\bf k}}
\Delta\sigma_{{\bf k}-{\bf k'}}}=\delta({\bf q}-{\bf k'})\times
\overline{\Delta\sigma_{{\bf q}-{\bf k}}\Delta\sigma_{{\bf k}-{\bf q}}}$.

Thus, for averaged Green function in the momentum space we have finally

\begin{equation}  \label{5} 
\overline{\ll S^+_{\bf q}|S^-_{\bf q'}\gg }=\delta({\bf q}+{\bf q'})
{{2\sigma +C(q,E)}\over{E-E_0(q)-\Sigma(q,E)}}
\end{equation}
where
\begin{eqnarray}
C(q,E)={1\over V}\sum_{\bf k}{{(J(|{\bf q}-{\bf k}|)-J(k))}
\over{(E-E_0(k))}}\overline{\Delta\sigma_{{\bf q}-{\bf k}}
\Delta\sigma_{{\bf k}-{\bf q}}}
\nonumber \\
\Sigma(q,E)={N\over V^2}\sum_{\bf k}{{(J(|{\bf q}-{\bf k}|)-J(q))
(J(|{\bf q}-{\bf k}|)-J(k))}
\over{E-E_0(k)}}\overline{\Delta\sigma_{{\bf q}-{\bf k}}
\Delta\sigma_{{\bf k}-{\bf q}}}
\nonumber
\end{eqnarray}.

\bigskip\bigskip\bigskip
\setcounter{equation}{0}
\renewcommand{\theequation}{2.\arabic{equation}}
\centerline{ \Large \bf 2. The relation between atomic momenta fluctuations}
\nopagebreak\bigskip\nopagebreak
\centerline{ \Large \bf  and density fluctuations.}
\nopagebreak\bigskip\nopagebreak

We start from the equation for mean atomic momenta $\sigma_i$

\begin{equation}  \label{7} 
\sigma_i={1\over 2}\tanh{\left \{ {1\over {2T}}\left (
\sum_j J(R_{ij})\Theta(R_{ij}-a)\sigma_j+h\right )\right \}}.
\end{equation}
Note, that
\begin{equation}  \label{8} 
\sum_j J(R_{ij})\Theta(R_{ij}-a)\sigma_j={N\over V}J(0)\sigma+
{N\over V}\sum_{\bf k}J(k) e^{i{\bf k}{\bf R}_j}
{1\over\sqrt{N}}\Delta\sigma_{\bf k}
\end{equation}

In linear approximation over $\Delta\sigma_{\bf q}$ using (\ref{7}) and
(\ref{8}) we obtained:
\begin{equation}  \label{9} 
\Delta\sigma_{\bf q}={\sigma\over{1-\left ({1\over 4}-\sigma^2\right )
{N\over V}J(q){1\over T}}}\rho_{\bf q}
\end{equation}
where $\rho_{\bf q}={1\over\sqrt{N}}\sum\limits_j e^{-i{\bf k}{\bf R}_j}
-\sqrt{N}\delta({\bf q})$
is the Fourier transform of the atomic density fluctuations, $T$ is the
temperature, $\sigma$ satisfies the equation

\begin{equation}  \label{10} 
\sigma={1\over 2}\tanh{\left \{ {1\over {2T}}\left (
{N\over V}J(0)\sigma+h\right )\right \}}.
\end{equation}

Now, using (\ref{9}) we can evaluate

\begin{equation}  \label{11} 
\overline{\Delta\sigma_{{\bf q}-{\bf k}}\Delta\sigma_{{\bf k}-{\bf q}}}=
\sigma^2\tilde{S}(|{\bf k}-{\bf q}|)=
{\sigma^2\over{\left \{1-\left ({1\over 4}-\sigma^2\right )
{N\over V}J(|{\bf k}-{\bf q}|){1\over T}\right \}}^2}S(|{\bf k}-{\bf q}|)
\end{equation}
where $S(q)=\overline{\rho_{\bf q}\rho_{\bf -q}}$ is the structure factor of
the amorphous material.

\bigskip\bigskip\bigskip
\setcounter{equation}{0}
\renewcommand{\theequation}{3.\arabic{equation}}
\centerline{ \Large \bf 3. Energy spectrum and damping of spin waves} \bigskip

The equation for energy spectrum of spin waves can be obtained from the pole
of averaged Green function (\ref{5})

\begin{equation}  \label{12} 
E=E_0(q)+
{N\over V^2}\sum_{\bf k}{{(J(|{\bf q}-{\bf q}|)-J(k))
(J(|{\bf q}-{\bf k}|)-J(k))}
\over{E-E_0(k)}}
\sigma^2\tilde{S}(|{\bf k}-{\bf q}|)
\end{equation}

Extracting the damping in a standard way we have

\begin{equation}  \label{13} 
\Gamma=
{N\over V^2}\sum_{\bf k}{(J(|{\bf q}-{\bf q}|)-J(k))
(J(|{\bf q}-{\bf k}|)-J(k))}
\sigma^2\tilde{S}(|{\bf k}-{\bf q}|)\delta(E-E_0(k))
\end{equation}

Beside the nonregularity in atoms' displacement equations (\ref{12})
and (\ref{13}) take
into account the structural fluctuations of mean atomic momenta that is
involved in the denominator of right hand side of (\ref{11}).

The long-wave asymptotic of energy spectrum is given by

\begin{equation}  \label{14} 
E=h+Dq^2,\qquad q\to 0
\end{equation}
where $D$ is the spin-wave stiffness constant

\begin{eqnarray}
\label{15} 				
D = D_0+\Delta\tilde{D},\qquad
 D_0=-\sigma{N\over V}J'(0),  \\
\label{17} 				
\Delta\tilde{D}=-\sigma{1\over{2\pi^2}}\int\limits_0^\infty dk
\thinspace k^2 \tilde{S}(k)
\left\{J'(k)+{2\over 3}k^2 J''(k)+{4\over 3}k^2{{(J'(k))^2}\over{J(0)-J(k)}}
\right\}, \\
{\rm where}\qquad J'(k)={{\partial J(k)}\over{\partial k^2}},\qquad
J''(k)={\partial\over{\partial k^2}}{\partial\over{\partial k^2}}J(k).
\nonumber
\end{eqnarray}

Let us assume that $J(k)$ is localized in the vicinity of zero.
Then we can perform the following approximation

\begin{equation}  \label{18} 		
\tilde{S}(k)\to{1\over{\left \{1-\left ({1\over 4}-\sigma^2\right )
{N\over V}J(0){1\over T}\right \}}^2}S(k)
\end{equation}

Substituting (\ref{18}) into (\ref{17}) we obtain

\begin{equation}  \label{19} 		
D=D_0+{1\over{\left \{1-\left ({1\over 4}-\sigma^2\right )
{N\over V}J(0){1\over T}\right \}}^2}\Delta D,
\end{equation}
where $\Delta D$
is given by (\ref{17}) after $\tilde{S}(k)$ is changed to $S(k)$.

As it follows from (\ref{19}), the spin-wave stiffness constant will increase
while applied field increases in the case $\Delta D<0$. Thus, the taking into
account the structure fluctuations allows to explain the anomalous behaviour
of spin-wave stiffness constant in amorphous ferromagnets.

Let us write the expression for the long-wave asymptotic of damping
$(q\to 0)$

\begin{equation}  \label{20} 		
\Gamma={7\over 12\pi}{V\over N}D{S(0)\over{\left \{1-\left
({1\over 4}-\sigma^2\right )
{N\over V}J(0){1\over T}\right \}}^2}q^5,
\end{equation}

Beside the scattering on the structure inhomogeneities spin excitations
scatter on the spin momenta fluctuations as well, this fact leads to
increase of damping. Momenta fluctuations disappear for $T\to 0$ or
$h\to\infty$, then $\sigma\to 1/2$ and $\tilde{S}(k)\to S(k)$.
That is why the contribution to damping at $T=0$ comes only for the
scattering processes of excitations on inhomogeneities of structure.

\end{document}